\documentclass[prl,aps,twocolumn,showpacs,preprintnumbers,amsmath,amssymb,
superscriptaddress]{revtex4}

\usepackage{graphicx}
\usepackage{dcolumn}
\usepackage{bm}

\font\msytw=msbm9 scaled\magstep1

\let\a=\alpha     \let\d=\delta 
  \let\h=\eta    \let\k=\kappa \let\l=\lambda
\let\m=\mu    \let\n=\nu    \let\x=\xi     \let\p=\pi    \let\r=\rho
\let\s=\sigma    \let\f=\varphi 
   \let\o=\omega
\let\G=\Gamma \let\D=\Delta  \let\L=\Lambda 
         
\let\O=\Omega 

 \def\VV{{\cal V}}
 
 \def\BBB{{\cal B}}

   \def\pp{{\bf p}}
 \def\xx{{\bf x}} \def\yy{{\bf y}} 

\def\kk{{\bf k}}

\def\RRR{\hbox{\msytw R}}

\def\\{\hfill\break}
\let\==\equiv
\let\io=\infty
\let\0=\noindent
\def\media#1{{\langle#1\rangle}}

\def\const{{\rm const}}
\def\tende#1{\,\vtop{\ialign{##\crcr\rightarrowfill\crcr\noalign{\kern-1pt
    \nointerlineskip} \hskip3.pt${\scriptstyle #1}$\hskip3.pt\crcr}}\,}
\def\otto{\,{\kern-1.truept\leftarrow\kern-5.truept\to\kern-1.truept}\,}

\def\to{\rightarrow}

\def\qed{\hfill\raise1pt\hbox{\vrule height5pt width5pt depth0pt}}

\def\V#1{{\bf#1}}
\def\be{\begin{equation}}
\def\ee{\end{equation}}
\def\bea{\begin{eqnarray}}
\def\eea{\end{eqnarray}}
\def\nn{\nonumber}
\def\pref#1{(\ref{#1})}

\begin{document}

\title{Lattice gauge theory model for graphene}
\author{Alessandro Giuliani}
\affiliation{%
Universit\`a di Roma Tre,
L.go S. L. Murialdo 1, 00146 Roma, Italy}
\author{Vieri Mastropietro}%
\affiliation{%
Universit\`a di Roma Tor Vergata,
Viale della Ricerca Scientifica 00133 Roma, Italy}
\author{Marcello Porta}
\affiliation{%
Universit\`a di Roma La Sapienza,
P.le A. Moro 2, 00185 Roma, Italy}

\begin{abstract}
The effects of the electromagnetic (e.m.) electron-electron
interactions in half-filled graphene are investigated in terms of
a lattice gauge theory model. By using exact Renormalization Group
methods and lattice Ward Identities, we show that the e.m.
interactions amplify the responses to the excitonic pairings
associated to a Kekul\'e distortion and to a charge density wave.
The effect of the electronic repulsion on the Peierls-Kekul\'e
instability, usually neglected, is evaluated by deriving an exact
non-BCS gap equation, from which we find evidence that strong e.m.
interactions among electrons facilitate the spontaneous distortion
of the lattice and the opening of a gap.
\end{abstract}

\maketitle Graphene, a monocrystalline graphitic film that has
been recently experimentally realized \cite{N1}, has highly
unusual electronic properties, due to the quasi-relativistic
nature of its charge carriers \cite{C}. In the presence of
long-ranged interactions, graphene provides an ideal laboratory
for simulating Quantum Field Theory (QFT) models at low energies
and to possibly observe phenomena like spontaneous chiral symmetry
breaking and mass generation. For a few years, there was
essentially no experimental signature of electron-electron
interactions in graphene, due to substrate-induced perturbations
that obscured their effects. However, the realization of suspended
graphene samples is allowing people to collect increasing evidence
of interaction effects \cite{A}. On the theoretical side,
understanding the properties of a system of interacting fermions
on the honeycomb lattice is a challenging problem, similar to
quantum electrodynamics, but with some peculiar differences that
make its study new and non-trivial \cite{Se}; its comprehension is essential
for graphene and, at the same time, it has relevance for other
planar condensed matter systems, like high $T_c$ superconductors,
and even for basic questions in QFT.

Most theoretical analyses deal with a simplified effective
continuum model of massless Dirac fermions with static Coulomb
interactions. Early results based on lowest order perturbation
theory predicted a {\it growth} of the effective Fermi velocity
$v(\kk)$ close to the Fermi points and excluded the spontaneous
formation of a gap at weak coupling \cite{V3}. The absence of a
gap is a serious drawback for possible technological applications
of graphene. Therefore, people started to investigate possible
mechanisms for its generation. One way to induce a gap is by
adding an interaction with an external periodic field, which can
be generated, e.g., by the presence of a substrate \cite{Z}. In
the absence of a substrate, Ref.\cite{K} proposed a mechanism
similar to the one at the basis of spontaneous chiral symmetry
breaking in strongly-coupled QED$_3$; the applicability of these
proposals to real graphene is a delicate issue, due to the
uncontrolled approximations related to a large-N expansions.
Another possible mechanism for gap generation is based on a
Peierls-Kekul\'e distortion of the honeycomb lattice, which is a
prerequisite for electron fractionalization \cite{J,J1}. One
optimizes over the distortion pattern, by minimizing the
corresponding electronic energy. In the absence of interactions, a
rather strong interaction with the classical phonon field is
needed for the formation of a non-trivial distortion \cite{J},
while the effects of the electron interactions is still poorly
understood.

In this Communication, we investigate the effects of electronic
interactions in terms of a lattice gauge theory model. We show
that the electronic interactions amplify the response functions
associated to Kekul\'e (K) or Charge Density Wave (CDW) pairings.
Moreover, we derive the exact form of the Peierls-Kekul\'e gap
equation in the presence of the e.m. electron-electron
interactions, from which we find evidence that strong e.m.
interactions enhance the Peierls-Kekul\'e instability (despite the
growth of the Fermi velocity \cite{S}, which apparently opposes
this effect). The model is analyzed by the methods of {\it
Constructive QFT}, which have already proved effective in
obtaining rigorous non-perturbative results in many similar
problems \cite{GM}; we rely neither on the effective Dirac
description nor on a large-N expansion.

\begin{figure}[htbp]
\centering
\includegraphics[width=0.25\textwidth]{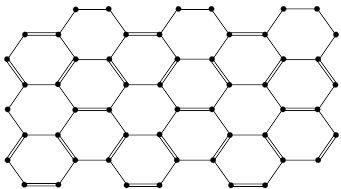}
\caption{The Kekul\' e pattern; the hopping parameter is $t +
\D_0$ and $t - \D_0/2$ on the double and single bonds respectively.}
\label{fig2}
\end{figure}

The charge carries in graphene are described by tight binding
electrons on a honeycomb lattice coupled to a three-dimensional
(3D) quantum e.m. field. We introduce creation and annihilation
fermionic operators $\psi_{\vec x,\s}^\pm=(a^\pm_{\vec x,\s},
b^\pm_{\vec x + \vec \d_1,\s})= |\BBB|^{-1}\int_{\vec
k\in\BBB}d\vec k\,\psi^\pm_{\vec k,\s} e^{\pm i\vec k\vec x}$ for
electrons with spin index $\s=\uparrow\downarrow$ sitting at the
sites of the two triangular sublattices $\L_A$ and $\L_B$ of a
honeycomb lattice; we assume that $\L_A=\L$ has basis vectors
$\vec l_{1,2}= \frac12(3,\pm\sqrt3)$ and that
$\L_B=\L_A+\vec\d_j$, with $\vec \d_1=(1,0)$ and
$\vec\d_{2,3}=\frac12 (-1,\pm\sqrt3)$ the nearest neighbor
vectors; $\BBB$ is the first Brillouin zone. The honeycomb lattice
is embedded in $\RRR^3$ and belongs to the plane $x_3=0$. The
grand-canonical Hamiltonian at half-filling is $H=H_0+H_C+H_A$,
where
\be
H_0=-t\sum_{\vec x\in\L_A} \sum_{j,\s} a^{+}_{\vec x,\s}
b^{-}_{\vec x + \vec \d_j,\s} e^{ie\int_0^1\vec\d_j\cdot\vec
A(\vec x+s\vec\d_j,0)\,ds}
 + c. c.
\label{aa} \ee
with $t$ the hopping parameter and $e$ the electric charge;
the coupling with the e.m. field
is obtained via the Peierls substitution. Moreover, if $n_{\vec
x}=\sum_\s a^+_{\vec x,\s}a^-_{\vec x,\s}$ (resp. $n_{\vec
x}=\sum_\s b^+_{\vec x,\s}b^-_{\vec x,\s}$) for $\vec x\in \L_A$
(resp. $\vec x\in\L_B$),
$$H_C=\frac{e^2}2\sum_{\vec x,\vec y\in\L_A\cup\L_B}
(n_{\vec x}-1)\f({\vec x}-{\vec y})(n_{\vec y}-1)\;,$$
where $\hat\f_{\vec p}$
is an ultraviolet regularized version of the static Coulomb potential.
Finally, $H_A$ is the energy (in the presence of an ultraviolet
cutoff) of the 3D photon field $\underline A=(\vec
A,A^3)$ in the Coulomb gauge. We fix units so that the speed of
light $c=1$ and the free Fermi velocity $v=\frac32t\ll1$. If we
allow distortions of the honeycomb lattice, the hopping becomes a
function of the bond length $\ell_{\vec x,j}$ that, for small
deformations, can be approximated by the linear function $t_{\vec
x,j}=t+\phi_{\vec x,j}$, with $\phi_{\vec x,j}=g( \ell_{\vec x,j}-\bar\ell)$
and $\bar\ell$ the
equilibrium length of the bonds. The Kekul\' e dimerization pattern
corresponds to, see Fig.\ref{fig2},
\be \phi_{\vec x,j}=\phi_0+\D_0\cos\big(\vec p_F^+(\vec \d_j-\vec
\d_{j_0}-\vec x)\big)\;,\label{ava1}\ee
with $j_0\in\{1,2,3\}$. In order to investigate the effect of the
e.m. interactions and the Peierls-Kekul\'e instability, we use the following
strategy: we first compute the ground state
energy and the correlations for $\phi_{\vec x,j}=0$ by exact
Renormalization Group (RG) methods \cite{BG}, finding, in
agreement with previous analyses \cite{V3b}, that the
quasi-particle weight vanishes at the Fermi points and the
effective Fermi velocity tends to the speed of light as power laws
with non-universal critical exponents. In addition, the analysis
of the response functions and of the corresponding exponents
indicates a tendency towards excitonic pairing;
the mass terms of K or CDW type are strongly amplified by the
interactions. Next, we compute the electronic correlations in the presence of
a non-trivial lattice distortion $\phi_{\vec x,j}$
and we show that a Kekul\' e dimerization pattern Eq.\pref{ava1} is a
stationary point of the total
energy (i.e., the sum of the elastic energy and the electronic
energy in the Born-Oppenheimer approximation).

\begin{figure}[htbp]
\centering
\includegraphics[width=0.33\textwidth]{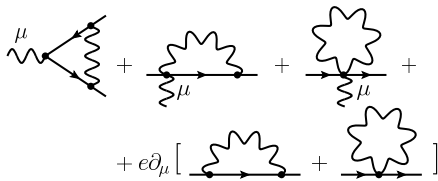}
\caption{The second order graphs contributing to the {\it dressed
charge} $e_{-\io}$, given by the contribution of the vertex part
minus the graphs coming from the wave function renormalization (if
$\m=0$) or the velocity renormalization ($\m=1,2$); their sum is
exactly vanishing, in agreement with the WI Eq.\pref{por1}.
} \label{fig1}\end{figure}

We start with $\phi_{\vec x,j}=0$; the analysis is very
similar to the one performed in the continuum Dirac approximation
in \cite{GMP} (which we refer to for more details), the main
difference being that the present lattice gauge theory model is
automatically ultraviolet-finite and gauge invariant: this avoids the
need for an ultraviolet regularization, which can lead to well known
ambiguities \cite{M}. The $n$-points imaginary-time correlations can be
obtained by the generating functional
\be e^{W(J,\l)}=\int P(d\psi)\int
P(dA)e^{\VV(A+J,\psi)+(\psi,\l)}\label{por} \ee
where: $\psi^{\pm}_{\kk}$ are Grassmann variables, with $\kk = (k_0,\vec k)$ 
and $k_0$ the Matsubara frequency, $P(d\psi)$ is the fermionic gaussian integration with inverse
propagator \be g^{-1}(\kk)= -\frac{1}{Z}\left(\begin{array}{cc}
ik_0 & v \O^*(\vec k)
\\ v \O(\vec k) & ik_0 \end{array}\right)\;\label{vo}\ee with $Z=1$
and $\O(\vec k) = \frac23\sum_{j=1,2,3}e^{i\vec k(\vec \d_j -
\vec\d_1)}$ (note that $g(\kk)$ is singular only at the Fermi
points $\kk=\kk_F^{\pm}
=(0,\frac{2\p}{3},\pm\frac{2\p}{3\sqrt3})$); if $\m=0,1,2$,
$A^\m(\pp)$ are gaussian variables with propagator $w_{\m\n}(\pp)=
\d_{\m\n}\int \frac{dp_3}{(2\pi)}\frac{\chi(|\vec
p|^2+p_3^2)}{\pp^2+p_3^2}$, where $\chi$ is an ultraviolet cutoff function; 
finally $\VV=e Z \int [j_0 A_0+v \vec
j \vec A]+h.o.t.$, where {\it h.o.t.} indicates higher order
interaction terms in $A$ produced by the Taylor expansion of the
exponential in $H_0$ and $j_\m$ is the bare lattice current
\footnote{$j_\m=\int
\frac{d\kk}{2\p|\BBB|}\,\frac{d\pp}{(2\pi)^3}\, \psi_{\kk+\pp}^{+}
\G_{\m}(\kk,\pp)\psi^{-}_{\kk}$ where $\G_{\m}$ are matrices which
can be deduced from Eq.\pref{aa}, with $\G_0(\kk_F^\o,0)=-i I$,
$\G_1(\kk_F^\o,0)=-\s_2$, $\G_2(\kk_F^\o,0)=-\o\s_1$}. We compute
$W(J,\l)$ via a rigorous Wilsonian RG scheme, writing the fields
$\psi,A$ as sums of fields $\psi^{(k)},A^{(k)}$, living on
momentum scales $|\kk-\kk_F^\pm|,|\pp|\simeq M^{k}$, with $k\le 0$
a scale label and $M>1$ a scaling parameter; the iterative
integration of the fields on scales $h<k\le 0$ leads to an
effective theory similar to \pref{por} with an ultraviolet cut-off
around the Fermi points of width $M^h$ and with effective scale
dependent wave function renormalization $Z_h$, Fermi velocity
$v_h$ and effective charge $e_h$.
This approach works only if $e_h$ does not flow to strong
coupling; the boundedness of $e_h$ follows from an exact Ward Identity (WI),
derived by the lattice phase transformation $\psi^\pm_{\xx,\s}\to
e^{\pm i e \a_\xx}\psi^\pm_{\xx,\s}$ in $W_h(0,\l)$
\footnote{$W_h$ is defined in the same way as $W$, with an extra
infrared cutoff suppressing momenta smaller than $M^h$}:
\be\pp_\m \L_{\m}^{(h)}(\kk,\pp)=e\big[S^{-1}_{\kk+\pp}\G_0(\vec
k,\vec p) -\G_0(\vec k,\vec p)S^{-1}_{\kk}\big]\label{por1}\ee
where $S_{\kk}$ is the interacting propagator, $\G_0(\vec k,\vec p)=\begin{pmatrix} -i & 0\\
0& -i e^{-i\vec p\vec\d_1}\end{pmatrix}$ and
$\L_{\m}^{(h)}(\kk,\pp)$ is the vertex function \footnote{More
precisely, $\pp_\m [S_{\kk+\pp}\L_{\m}^{(h)}(\kk,\pp)S_{\kk}]_{ij}
=\frac{\partial}{\partial\a_\pp}\frac{\partial^2}
{\partial\l^-_{\kk,j}\partial\l^+_{\kk+\pp,i}} W_{h}(\partial
\a,\l)\big|_{\a=\l=0}$.} that, if computed at external momenta
$|\kk-\kk_{F}^{\pm}| \sim M^h$ and $|\pp|\ll M^h$, is proportional to $e_{h}$ (if
$\m=0$) or $e_h v_h$ (if $\m=1,2$). Using Eq.(\ref{por1}), we find
that $e_{h}\to e_{-\io}= e+e^3 F(e)$, with $F(e)$ a series in
$e^2$ with bounded coefficients, i.e., the effective charge tends
to a line of fixed points; $F(e)$ is vanishing at lowest order,
see Fig.\ref{fig1}, but the WI does not exclude that $F(e)$ is non-zero at
higher orders. A similar WI implies that the photon remain
massless and,
as an outcome of the above procedure, we get an expansion of the
Schwinger functions in the effective couplings $e_h$ that is {\it
finite at all orders}, see \cite{GMP} for the proof; this is in
contrast with the naive perturbation theory, which is plagued by
logarithmic divergencies. The boundedness of $e_h$ makes such
expansion meaningful and it allows one to control the flow of
$Z_h$ and $v_h$. One finds that: (i) $\lim_{h\to-\io}v_h=1$, i.e.,
Lorentz invariance spontaneously emerges; (ii) both $Z_h^{-1}$ and
$1-v_h$ vanish with two anomalous power laws, see \cite{GMP}.
Therefore, if $\phi_{\vec x,j}=0$, the dressed propagator has a
form similar to Eq.\pref{vo}, with $Z$ and $v$ replaced by
$Z(\kk-\kk_{F}^{\pm})$ and $v(\kk-\kk_{F}^{\pm})$; if $\kk$ is far from the Fermi points,
$Z(\kk-\kk_{F}^{\pm})$ and $v(\kk-\kk_{F}^{\pm})$ are close to their unperturbed values,
namely $1$ and $\frac32t$. On the contrary, if $|\kk'|\ll 1$, $Z(\kk') \sim |\kk'|^{-\h}$, with
$\h=\frac{e^{2}}{12\pi^2} +\cdots$ an anomalous critical exponent
that is finite at all orders in $e^2$, and $v(\kk')$ tends to the
speed of light. Moreover, $1-v(\kk')\sim (1-v)|\kk'|^{\tilde
\h}$, with $\tilde\h= \frac{2e^{2}}{5\pi^2 } +\cdots$ another
anomalous critical exponent.

The above analysis confirms, at all orders and in the presence of a
lattice cut-off, the results found long ago in \cite{V3b}, where
graphene was described by an effective continuum Dirac model with
an ultraviolet {\it dimensional} regularization; the exponents
agree at lowest order. Gauge invariance implies that the dressed
Fermi velocity has a universal value (the speed of light) and
Lorentz invariance emerges; this is what happens both in \cite{V3b}
(thanks to the use of dimensional regularization) and in the
present more realistic lattice model, thanks to the exact lattice
WI Eq.\pref{por1}. On the other hand, we expect that as soon as
gauge invariance is broken the limiting Fermi velocity is smaller
than the speed of light; this is indeed what happens in
\cite{GMP}, where gauge invariance is broken by the momentum
cut-off.

In order to understand which instabilities are likely to occur in
the system, we compute response functions
or generalized susceptibilities (which apparently have never been
systematically computed before, not even in the Dirac approximation);
more precisely, for $\phi_{\vec x,j}=0$, we compute
$R^{(\a)}(\xx,j;\yy,j')=\media{\r^{(\a)}_{\xx,j}\r^{(\a)}_{\yy,j'}}$
with, e.g.,
$\r^{(K)}_{\xx,j}=\sum_\s\big(a^+_{\xx,\s}b^-_{\xx+(0,\vec\d_j),\s}
e^{ie\int_0^1\vec\d_j\cdot\vec A(\xx + s(0,\vec\d_j),0)\,ds}
+c.c.\big)$ or $\r^{(CDW)}_{\xx,j}=\sum_\s
\big(a^+_{\xx,\s}a^-_{\xx,\s}-
b^+_{\xx+(0,\vec\d_j),\s}b^-_{\xx+(0,\vec\d_j),\s}\big)$. The two latter
operators describe,
respectively, {\it inter-node} and {\it intra-node} excitonic
pairings of K and CDW type;
this is because the possible presence of a condensate in the
$\kk=\kk_F^{\pm}$ channel for
$\r^{(K)}_{\xx,j}$ or in the $\kk=\V0$ channel for $\r^{(CDW)}_{\xx,j}$
would signal the emergence of Long Range Order (LRO)
of K-type (see Fig.\ref{fig2}) or of CDW-type (a period-2 alternation of
excess/deficit of electrons in the sites of the A/B lattice).
Other relevant bilinears are the
Cooper pairings, i.e., linear combinations of terms of the form
$a^+_{\xx,\s}a^+_{\xx,-\s}$ or $a^+_{\xx,\s} b^+_{\xx+(0,\vec
\d_j),\s'}$. The large distances asymptotic behavior of the response functions
is:
$$R^{(\a)}(\xx,j;\V0,j)\sim G^{(\a)}_1(\xx)+\cos(\vec p_{F}^+\cdot\vec x)
G^{(\a)}_2(\xx)\;,$$
with $|G^{(\a)}_i(\xx)|\sim(\const.)|\xx|^{-\x^{(\a)}_i}$ two scaling invariant functions 
(similar formulas are valid for $j\neq j'$).
In the absence of interactions, $\x^{(\a)}_i=4$, for all $\a$ and
$i$; there are no preferred instabilities. The presence of the
interaction with the e.m. field removes the degeneracy in the
decay exponents: some responses are enhanced and some other
depressed. It turns out that \be \x^{(CDW)}_{1}=4-4e^2/(3\pi^2)+
\cdots,\quad\x^{(K)}_2= 4-4e^2/(3\pi^2)+\cdots\nn \ee On the
contrary, $\x^{(CDW)}_{2}$ and $\x^{(K)}_1$ are vanishing at
second order, while all the Cooper pairs responses decay faster
than $|\xx|^{-4}$. The conclusion is that the responses to
excitonic pairing of K or CDW type are amplified by the e.m. interaction:
in this sense, we can say that the e.m. interaction
induces quasi-LRO of K and CDW type.

Correspondingly, possible small distortions or inhomogeneities of
the K or CDW type are dramatically enhanced by the interactions.
For instance, let us choose $\phi_{\vec x,j}$ as in
Eq.\pref{ava1}. The RG analysis can be repeated in the presence of
a Kekul\'e mass term $\D_0\sum_{\vec x,j} \cos\big(\vec p_F^+\cdot(\vec
x-\vec\d_j+\vec\d_{j_0})\big)\r^{(K)}_{\vec x,j}$ in the
Hamiltonian, which produces a new relevant coupling constant, the
effective Kekul\'e mass.
Therefore, the interaction produces an effective
momentum-dependent gap $\D(\kk)$ that increases with a power law
with exponent $\h^K$ from the value $\D_0$ up to \be
\D(\kk_F^\pm)= \D_0^{1/(1+\h^K)}\;,\quad \h^K=2e^2/(3\p^2)+\cdots
\ee Note that the ratio of the dressed and bare gaps {\it
diverges} as $\D_0\to 0$. The enhancement of the dressed gap is
related to the phenomenon of gap generation in \cite{K}, but it is
found here avoiding any unrealistic large-$N$ expansion. A similar
enhancement is found for the gap due to a CDW modulation.

Finally, let us discuss a possible mechanism for the
spontaneous distortion of the lattice and the opening of a gap
(Peierls-Kekul\'e instability). We use a variational
argument, which shows that a Kekul\' e dimerization pattern of the form
Eq.\pref{ava1} is a stationary point of the total energy
$\frac{\k}{2g^2}\sum_{\vec x,j}\phi^2_{\vec x,j}+ E_0(\{\phi_{\vec x,j}\})$,
where the first term is the elastic energy and $E_0(\{\phi_{\vec x,j}\}$
is the electronic ground state energy in the Born-Oppenheimer approximation.
The extremality condition for the energy is $\k\phi_{\vec x,j}=g^2
\media{\r^{(K)}_{\vec x,j}}^{\phi}$, where $\media{\cdot}^{\phi}$
is the ground state average in the presence of the distortion pattern
$\{\phi_{\vec x,j}\}$.
Computing $\media{\r^{(K)}_{\vec x,j}}^{\phi}$ by RG with the
multiscale analysis explained above, we find that Eq.(\ref{ava1})
is a stationary point of the total energy, provided that
$\phi_0=c_0 g^2/\k+\cdots$ for a suitable constant $c_0$ and that
$\D_0$ satisfies the following non-BCS gap equation:
$$\D_0\simeq \frac{g^2}{\k}\!\!\!\int\limits_{\D\lesssim|\kk'|\lesssim 1}
\!\!\!\!\!\!\! d\kk' \frac{ Z^{-1}(\kk')\D(\kk')|\O(\vec k')|^2}
{k_0^2+v^2(\kk')|\O(\vec k'+ \vec p_F^\o)|^2+|\D(\kk')|^2}$$
where $\D=\D_0^{1/(1+\h^K)}$ and, for $\D\lesssim|\kk'| \ll1$,
$Z(\kk') \sim |\kk'|^{-\h}$, $1-v(\kk')\sim
(1-v)|\kk'|^{\tilde\h}$ and $\D(\kk')\simeq \D_0\,
|\kk'|^{-\h^K}$.
In the absence of interactions, $Z(\kk')=1$, $v(\kk')=v$ and the
above equation reduces to the free one in \cite{J}. Our gap
equation is qualitatively equivalent to the simpler expression
\be 1=g^{2}\int_{\D}^{1}d\r\,\frac{\r^{\eta - \eta_K}}{1 - (1 -
v)\r^{\tilde\eta}} \label{sf}\ee from which its main features can
be easily inferred.

At weak e.m. coupling, the integral in the right hand side
(r.h.s.) is infrared {\it convergent}, which implies that a non
trivial solution is found only for $g$ larger than a critical
coupling $g_c$; remarkably, $g_c\sim\sqrt{v}$, with $v$ the free
Fermi velocity, even though the effective Fermi velocity flows to
the speed of light. Therefore, at weak coupling, the prediction
for $g_c$ is qualitatively the same as in the free case \cite{J};
this can be easily checked by noting that the denominator in the
r.h.s. of Eq.(\ref{sf}) is sensibly different from $v$ only if
$\r$ is exponentially small in $v/\tilde\h$.

On the other hand, if one trusts our gap equation also at {\it
strong} e.m. coupling and if in such a regime $\h^K-\h=
\frac{7e^2}{12\p^2}+\cdots$ exceeds $1$, then the r.h.s. of
Eq.\pref{sf} {\it diverges} as $\D\to 0$, {\it which guarantees
the existence of a non-trivial solution for arbitrarily small}
$g$; this can be easily checked by rewriting Eq.\pref{sf}, up to
smaller corrections, as $1\simeq \frac{g^2}{\h^K-\h-1}
\D^{1+\h-\h^K}$, that is $\D_0\simeq g^{2(1+\h^K)/(\h^K-\h-1)}$:
note the non-BCS form of the gap, similar to the one appearing in
certain Luttinger superconductors \cite{M1}. The existence of a
non trivial solution for arbitrarily small $g$ suggests that
strong e.m. interactions between fermions enforce the
Peierls-Kekul\'e mechanism and facilitate the spontaneous
distortion of the lattice and the gap generation, by lowering the
critical phonon coupling $g_c$; this is in agreement with the
one-dimensional case, where the Peierls instability is enhanced by
the electronic repulsion, see \cite{one}. Note the crucial role in
the above discussion played by the momentum dependence of the gap
term (leading to the factor $\r^{\eta - \eta_K}$ in Eq.\pref{sf}); on
the contrary, the growth of the Fermi velocity plays a minor role.
A similar analysis can be repeated for the gap generated by a CDW instability.

In conclusion, we considered a lattice gauge theory model for
graphene and we predicted that the electron repulsion enhances
dramatically, with a non-universal power law, the gaps due to the
Kekul\'e distortion or to a density asymmetry between the two
sublattices, as well as the responses to the corresponding
excitonic pairings. Moreover, we derived an exact non-BCS gap
equation for the Peierls-Kekul\'e instability from which we find
evidence that strong e.m. interactions facilitate the spontaneous
distortion of the lattice and the gap generation, by lowering the
critical phonon coupling.

A.G. and V.M. gratefully acknowledge financial support from the
ERC Starting Grant CoMBoS-239694. We thank D. Haldane and M.
Vozmediano for many valuable discussions.

\bibliographystyle{amsalpha}

\end{document}